\documentclass{article}

\usepackage{graphicx}

\begin{document}

\begin{center}
{\Large \bf Quantitative Evaluation of Decoherence and{}\\{}\vskip0.16cm{}Applications for
Quantum-Dot Charge Qubits}
\end{center}

\begin{center}{\bf Leonid Fedichkin}\ and\ {\bf Vladimir Privman}
\end{center}

\begin{center}{Center for Quantum Device Technology,
Department of Physics and{}\\{}Department of Electrical and Computer Engineering,
Clarkson University, Potsdam, New York 13699--5721, USA }\end{center}


\section*{Abstract}
We review results on evaluation of loss of information in quantum registers
due to their interactions with the environment. It is demonstrated that an
optimal measure of the level of quantum noise effects can be introduced via
the maximal absolute eigenvalue norm of deviation of the density matrix of
a quantum register from that of ideal, noiseless dynamics. For a semiconductor
quantum dot charge qubits interacting with acoustic phonons, explicit expressions
for this measure are derived. For a broad class of environmental modes, this measure
is shown to have the property that for small levels of quantum noise it is additive
and scales linearly with the size of the quantum register.

\section{Introduction}

In recent years, there has been significant progress in quantum
computation and design of solid-state quantum information
processors~\cite{Shor97,Grover97,Loss98,Privman98,Kane98,Vrijen,Nano00,Makhlin01,IV3,IV4,Vion02,Chiorescu03,Hanson03,Pashkin03,IV2,Shlimak04}.
Quantum computers promise enormous speed-up of
computation of certain very important problems, including
factorization of large numbers~\cite{Shor97} and search~\cite{Grover97}. However, practically
useful quantum information processing devices have not been made yet. One of the
major obstacles to scalability has been decoherence. This is due to the fact
that the effect of quantum speed-up is crucially dependent upon the
coherence of quantum registers. Therefore, understanding the dynamics of coherence loss has
drawn significant experimental and theoretical effort.

In general, decoherence~\cite{vanKampen,nonMarkov,Anastopoulos,Ford,Braun,Lewis,Wang,PMV,Privman,Lutz,Khaetskii,OConnell,Strunz,Haake,short,VP6,VPAF5,VPDT4,VPDS3,VPDT2,VPDT1,Solenov}
reveals itself in most experiments with quantum objects. It is a process whereby the
quantum coherent physical system of interest interacts with the environment and, because of this
interaction, changes its evolution from unperturbed ``ideal'' dynamics. The change of the
dynamics is reflected by the corresponding change of the density
matrix~\cite{Neumann,Abragam,therm2,therm1,Louisell} of the system.
The time-dependence of the system's density matrix should be
evaluated for an appropriate model of the system and its environment.
If a multi-particle quantum system is considered then the respective density matrix
becomes rather large and difficult to deal with. This occurs even for relatively small quantum
registers containing just a few quantum bits (qubits).
In this paper, we review evaluation
of decoherence effects starting from the system Hamiltonian and followed by
the definition and estimation of a decoherence error-measure in a quantum
information processing ``register'' composed of
several qubits.

The paper is organized as follows: In Section~\ref{Sec2}, we consider a specific
example of a solid state nanostructure.
As a representative model for a qubit, we consider an electron in a semiconductor double quantum
dot system. We derive the evolution of the density
matrix of the electron, which losses
coherence due to interaction with phonons. In Section~\ref{Sec3}, we define a
measure characterizing decoherence and show how to calculate it
from the density matrix elements for a semiconductor double quantum
dot system introduced earlier.
Finally, in Section~\ref{Sec4}, we establish that the measure of decoherence introduced,
is additive for several-qubit registers, i.e., the total ``computational error'' scales
linearly with the number of qubits.

\section{Semiconductor Quantum Dot Charge Qubit}\label{Sec2}

Solid-state nanostructures attracted much attention recently as a possible
basis for large scale quantum information processing~\cite{roadmap}.
Most stages of their fabrication can be borrowed from existing fabrication steps
in microelectronics industry. Also, only microelectronics technology has demonstrated the ability
to create and control locally evolution of thousands of nano-objects, which is required for
quantum computation. There were several proposals for semiconductor qubits, reviewed, e.g., in~\cite{PMV}.
In particular, the encoding of quantum information in the position of the electron was investigated
in~\cite{Barenco,Hawrylak,Alex1,Alex2,Openov}. In~\cite{Ekert}
it was argued that an electron in a typical quantum dot will loose coherence very fast
which will prevent it from being a good qubit. However, this problem can be
resolved with sophisticated designs of quantum-dot arrangements, e.g.,
arrays of several quantum dots, if properly designed~\cite{Zanardi}, can form a
coherent quantum register. It was also shown that a
symmetric layout of just two quantum dots can strongly diminish decoherence effects due
to phonons and other environmental noises~\cite{Fedichkin,dd,dd_ieee}.

Recent successful observations~\cite{Hayashi,Marcus,Fuji1,Fuji2,Fuji3} of
spatial evolution of an electron in symmetric semiconductor
double dot systems have experimentally confirmed that such a system
is capable of maintaining coherence at least on time scales sufficient for
observation of several cycles of quantum dynamics. In the above experiments measurements were performed
at very low substrate temperatures of few tens of mK, in order to avoid
additional thermally activated sources of decoherence.
Theoretical results
on the influence of the temperature on the first-order phonon relaxation rates in double
dot systems were presented in~\cite{Barrett,Ahn}.

In view of the above experimental advances,
we have chosen a single electron in semiconductor double quantum dot system,
whose dynamics is affected by
vibrations of the crystal lattice, as a representative example of a quantum
coherent system interacting with the environment.
In the range of parameters corresponding to
experiments~\cite{Hayashi,Fuji1,Fuji2,Fuji3} phonons dominate decoherence.
Of course, for different systems or for similar systems in different ranges of external conditions
some other sources of decoherence may prevail, for example, noise due to hopping of charge carries on
nearby traps, studied in~\cite{Pashkin,Martin}, or due to the electron-electron interaction~\cite{Vorojtsov}.

\begin{figure}[t]
\includegraphics[width=12 cm] {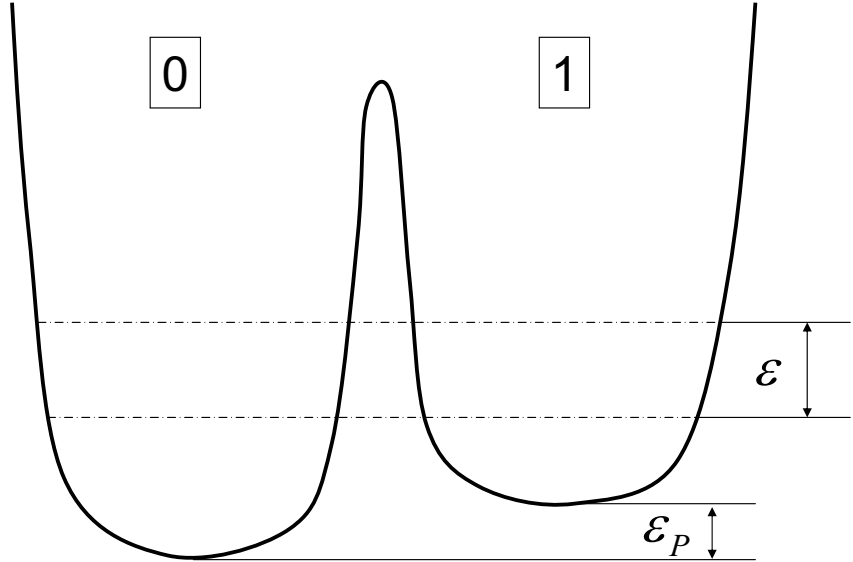}
\caption{Double well potential.} \label{fig:1}
\end{figure}

Semiconductor double quantum dot creates
three-dimensional double well confinement potential for electron in it. Let us denote the line connecting centers of the dots
as the $x$-axis. Then the electron confining potential along $x$, is schematically shown in Fig.~\ref{fig:1}.
The nanostructure is composed of two quantum dots with a potential barrier between them.
Parameters of the structure are properly adjusted so that two lower energy levels of spatial quantization
lie very close to each other compared to the external temperature and to the distances to higher energy levels.
Therefore hopping of the electron to higher levels is suppressed.
The electron is treated as a superposition of two basis states,
$|0\rangle$ and $|1\rangle$, corresponding to ``false'' and ``true'' in Boolean logic,
\begin{equation}
\psi  = \alpha \psi_0 + \beta \psi_1 .
\end{equation}
It should be noted that the states that define the ``logical'' basis are not the
ground and first excited states of the double-dot system.
Instead, $\psi_0$ (the ``0'' state
of the qubit) is chosen to be localized at the first quantum dot and, in a zeroth
order approximation, be similar to the ground state of that dot if it were
isolated. Similarly,  $\psi_1$ (the ``1'' state) resembles the ground state of
the second dot (if it were isolated). This assumes that the dots are sufficiently
(but not necessarily exactly) symmetric. We denote the coordinates of the
potential minima  of the dots (dot centers) as vectors $\mathbf R_0$ and $\mathbf
R_1$, respectively. The separation between the dot centers is
\begin{equation}
\mathbf L = \mathbf R_1-\mathbf R_0 .
\end{equation}

The Hamiltonian of an electron interacting with a phonon bath consists of three terms
\begin{equation} \label{H}
 H = H_e + H_p + H_{ep}.
\end{equation}
The electron term is
\begin{equation}\label{He}
 H_e = - \frac{1}{2}\varepsilon _{A}(t)\sigma _x- \frac{1}{2}\varepsilon _{P}(t)\sigma _z,
\end{equation}
where $ \sigma_x$ and $\sigma _z$ are Pauli matrices, whereas
 $\varepsilon _{A}(t)$ and $\varepsilon _{P}(t)$ can have time-dependence, as
 determined by unitary single-qubit quantum gate-functions to be implemented for
 specific quantum algorithm. This can be achieved by adjusting the potential on the
 metallic nanogates surrounding the double-dot system. For constant $\varepsilon _{A}$ and
$\varepsilon _{P}$, the energy splitting between the electron energy levels is
\begin{equation}
 \varepsilon=\sqrt{\varepsilon _{A}^2+\varepsilon _{P}^2}.
\end{equation}

The Hamiltonian term of the phonon bath is described by
\begin{equation}
 H_p = \sum\limits_{\mathbf{q},\lambda } \hbar \omega_q {\kern 1pt} b_{\mathbf{%
q},\lambda }^{\dagger}  b_{\mathbf{q},\lambda },
\end{equation}
where $ b^{\dagger}_{\mathbf{q},\lambda}$ and $ b_{\mathbf{q},\lambda}$ are
the creation and annihilation operators of phonons,
respectively, with
the wave vector $\mathbf q\,$ and polarization $\lambda$. We approximate the
acoustic phonon spectrum as isotropic one with a linear dispersion
\begin{equation}\omega_q = s q,\end{equation}
where $s$ is the speed of sound in the semiconductor crystal.

In the next few paragraphs we show that the electron-phonon interaction
can be expressed as
\begin{equation}  \label{int}
 H_{ep}=\sum\limits_{\mathbf{q},\lambda }
 \sigma_z {\kern 1pt}\left( g_{\mathbf{q},\lambda }  b_{\mathbf{q},
\lambda}^{\dagger} +  g_{\mathbf{q},\lambda }^* b_{\mathbf{q}, \lambda}\right),
\end{equation}
with the coupling constants $g_{\mathbf{{q}, \lambda}}$ determined by the
geometry of the double-dot and the properties of the material.
The derivation  follows \cite{dd,dd_ieee}. The piezoacoustic
electron-phonon interaction \cite{Mahan} is given by
\begin{equation}  \label{p1}
H_{ep}=i{\sum\limits_{\mathbf q,\lambda} }\sqrt{\frac \hbar {2\rho s q V
}}\, M_\lambda ({\mathbf q})F (\mathbf q)(b_{\mathbf
q}+b_{-\mathbf q}^{\dagger}),
\end{equation}
where $\rho$ is the density of the semiconductor, $V$ is volume of the sample,
 and for the matrix element
$M_\lambda ({\mathbf q})$, one can derive
\begin{equation}  \label{p2}
M_\lambda ({\mathbf q})=\frac 1{2q^2}{\sum\limits_{ijk} }(\xi
_i^{\vphantom{Q}}q_j^{\vphantom{Q}}+\xi
_j^{\vphantom{Q}}q_i^{\vphantom{Q}})q_k^{\vphantom{Q}} M_{ijk}.
\end{equation}
Here $\xi _j$ are the polarization vector components for
polarization $\lambda$, while $M_{ijk}$ express the electric field as a linear
response to the stress,
\begin{equation}  \label{p3}
E_k={\sum\limits_{ij} }M_{ijk}S_{ij}.
\end{equation}
For a crystal with zinc-blende lattice, like GaAs, the tensor $M_{ijk}$
has only those components non-zero for which all three indexes $i$, $j$, $k$ are
different; furthermore, all these components are equal, $M_{ijk}=M$. Thus, we have
\begin{equation}  \label{p4}
M_\lambda (\mathbf q)=\frac
M{q^2}(\xi^{\vphantom{Q}}_1q_2^{\vphantom{Q}}q_3^{\vphantom{Q}}+
\xi^{\vphantom{Q}}_2
q_1^{\vphantom{Q}}q_3^{\vphantom{Q}}+\xi^{\vphantom{Q}}_3q_1^{\vphantom{Q}}q_2^{\vphantom{Q}}).
\end{equation}

The form factor $F(\mathbf{{q})}$ accounting for that we are working with electrons which are
not usual plane waves, is given by
\begin{equation}\label{e-density}
F (\mathbf q)=
\sum\limits_{j,k}%
c_j^{\dagger}c_k\int d^3r \phi_j^{*}(\mathbf r)\phi_k(\mathbf r)e^{-i\mathbf
q\cdot \mathbf r},
\end{equation}
where $c_k$, $c^{\dagger }_j$ are the annihilation and creation operators of the
basis states $k,j=0,1$. In quantum dots formed by a repulsive potential of
nearby gates, an electron is usually confined near the potential minima, which
are approximately parabolic. Therefore the ground
states in each dot have Gaussian shape
\begin{equation}  \label{1}
\phi_j(\mathbf r)=\displaystyle\frac{e^{-|\mathbf r-\mathbf
R_j|^2/2a^2}}{a^{3/2}\pi ^{3/4}}\,,
\end{equation}
where $2a$ is a characteristic size of the dots.

We assume that the distance between the dots, $L = | \mathbf L |$, is sufficiently large compared
to $a$, and that the different dot wave functions do not strongly overlap,
\begin{equation}
\bigg| \int d^3r \phi_j^{*}(\mathbf r)\phi_k(\mathbf r)e^{-i \mathbf q\cdot
\mathbf r} \bigg| \ll 1, \quad {\rm for} \quad j\neq k.
\end{equation}
In other words tunneling between the dots is small, as
is the case for the recently studied experimental structures
\cite{Hayashi,Fujisawa,Dzurak1,Dzurak2}, where the splitting due to tunneling,
measured by $\varepsilon _{A}$, was just several tens of $\mu$eV, while the electron
quantization energy in each dot was at least several meV.

For $j=k$, we obtain
\begin{eqnarray}\label{3}\nonumber
\int d^3 r \phi_j^{*}(\mathbf r)\phi_j(\mathbf r) e^{-i\mathbf q\cdot \mathbf r}=
\frac{1}{a^3\pi^{3/2} }\int d^3 r e^{-|{\mathbf r} -{\mathbf
R}_j|^2/a^2}e^{-i\mathbf q\cdot \mathbf r}
\end{eqnarray}
\begin{eqnarray}
=e^{-i{{\mathbf q}\cdot {{\mathbf R}_j}}}e^{-a^2q^2/4}.
\end{eqnarray}
The resulting form factor is
\begin{equation}
F (q)=e^{-a^2q^2/4}e^{-i\mathbf q\cdot \mathbf R}(c_0^{\dagger }c_0e^{i\mathbf
q\cdot \mathbf L/2}+c_1^{\dagger }c_1e^{-i\mathbf q\cdot \mathbf L/2}),
\end{equation}
where $\mathbf R=\left(\mathbf R_0+\mathbf R_1\right)/2$.
Therefore
\begin{equation}\label{form}
F (q)=e^{-a^2q^2/4}e^{-i\mathbf q\cdot \mathbf R}\left[\cos (\mathbf q\cdot
\mathbf L /2)I+i\sin (\mathbf q\cdot \mathbf L /2)\sigma _z\right] \label{4},
\end{equation}
where $I$ is the identity operator. Only the last term in (\ref{form})
represents an interaction affecting the qubit states.
It leads to a Hamiltonian term of the form (\ref{int}), with coupling constants
\begin{eqnarray}  \label{p6}\nonumber
g_{\mathbf q, \lambda}&=&-\sqrt{\frac{\hbar   }{2 \rho q s V
}} \, M e^{-a^2q^2/4 - i\mathbf q\cdot \mathbf R}\\
&&\times(\xi_1^{\vphantom{Q}}e_2^{\vphantom{Q}}e_3^{\vphantom{Q}}+
\xi_2^{\vphantom{Q}}e_1^{\vphantom{Q}}e_3^{\vphantom{Q}}
+\xi_3^{\vphantom{Q}}e_1^{\vphantom{Q}}e_2^{\vphantom{Q}}) \sin (\mathbf q\cdot
\mathbf L/2),
\end{eqnarray}
 where $e_k=q_k/q$.

The general form of qubit evolution controlled by the Hamiltonian term (\ref{He})
is time dependent. Decoherence estimates for some solid-state systems with
certain shapes of time dependence of the system Hamiltonian were reported
recently \cite{Solenov,Brandes,Brandes2}. However, such estimations are rather
sophisticated. To avoid this difficulty we observe that all single-qubit rotations which
are required for quantum algorithms can be successfully performed by using two
constant-Hamiltonian gates without loss of quantum speed-up, e.g., by amplitude
rotation gate and phase shift gate
\cite{Kitaev3}. To implement these gates one can keep the Hamiltonian term
(\ref{He}) constant during the implementation of each gate, adjusting the
parameters $\varepsilon _{A}$ and $\varepsilon _{P}$ as appropriate for each gate
and for the idling qubit in between gate functions.
In the next paragraph we initiate our consideration of decoherence during the implementation of the NOT amplitude gate.
Then consider $\pi$-phase shift gate later in the section.

The quantum NOT gate is a unitary operator which transforms the states
$|0\rangle$ and $|1\rangle$ into each other. Any superposition of $|0\rangle$ and
$|1\rangle$ transforms accordingly,
\begin{equation}
    {\rm NOT} \left(\alpha |0\rangle + \beta |1\rangle\right) = \beta |0\rangle + \alpha |1\rangle .
\end{equation}
The NOT gate can be implemented by properly choosing $\varepsilon_A$ and
$\varepsilon_P$ in the Hamiltonian term (\ref{He}). Specifically, with constant
\begin{equation}
 \varepsilon_A=\varepsilon
\end{equation}
and
\begin{equation}
 \varepsilon_P=0,
\end{equation}
the ``ideal'' NOT gate function is carried out, with these interaction
parameters, over the time interval
\begin{equation}
 \tau=\frac{\pi\hbar}{\varepsilon}.
\end{equation}

The major source of quantum noise for double-dot qubit subject to the NOT-gate
type coupling, is relaxation involving energy exchange with the phonon bath
(i.e., emission and absorption of phonons). Here it is more convenient to
study the evolution of the density matrix in the energy basis, $\left\{
\left|+\right\rangle,\left|-\right\rangle\right\}$, where
\begin{equation}
\left|\pm\right\rangle=\left(\left|0\right\rangle \pm
\left|1\right\rangle\right)/\sqrt{2}.
\end{equation}
Then, assuming that the time interval of interest is $[0,\tau]$, the qubit
density matrix can be expressed \cite{therm2} in the energy basis as
\begin{eqnarray} \label{rho_rel}
\rho(t)=\left(
\begin{array}{cc}
  \rho_{++}^{th}+\left[\rho _{++}(0)-\rho_{++}^{th}\right]e^{ - \Gamma t} &
  \rho _{+-}(0)e^{ - (\Gamma/2-i\varepsilon/\hbar ) t} \\\\
  \rho _{-+}(0)e^{ - (\Gamma/2+i\varepsilon/\hbar) t} & \rho_{--}^{th}+\left[\rho _{--}(0)-\rho_{--}^{th}\right]e^{ - \Gamma t}  \\
\end{array}
\right)\! .
\end{eqnarray}
This is a standard Markovian approximation for the evolution of the density
matrix. For large times, this type of evolution would in principle
result in the thermal state, with
the off-diagonal density matrix elements decaying to zero, while the diagonal
ones approaching the thermal values proportional to the Boltzmann factors
corresponding to the energies $\pm \varepsilon /2$. However, here we are only
interested in such evolution for a relatively short time interval, $\tau$, of a NOT gate.
The rate parameter $\Gamma$ is simply the sum \cite{therm2} of the phonon emission rate,
$W^{e}$, and absorption rate, $W^{a}$,
\begin{equation}\label{Gamma}
    \Gamma=W^{e}+W^{a}.
\end{equation}

The probability for the absorption of a phonon due to excitation from the ground
state to the upper level is
\begin{equation}
  w^{\lambda}=\frac{2 \pi}{\hbar}|\langle f|H_{ep}|i\rangle|^2 \delta(\varepsilon
  -\hbar s q),
\end{equation}
where $|i\rangle$ is the initial  state with the extra phonon with energy $\hbar
s q$ and $|f\rangle$ is the final  state, $\mathbf q$ is the wave vector, and
$\lambda$ is the phonon polarization. Thus, we have to calculate
\begin{equation}\label{W}
W^a=\sum\limits_{\bf q, \lambda}
w^{\lambda}=\frac{V}{(2\pi)^3}\sum\limits_{\lambda}\int d^3 q\, w^{\lambda}.
\end{equation}
For the interaction (\ref{int}) one can derive
\begin{equation}
w^{\lambda}=\frac{2 \pi}{\hbar}|g_{\mathbf q,\lambda}|^2 N^{th}
\delta(\varepsilon-\hbar sq),
\end{equation}
where
\begin{equation}
N^{th}=\frac{1}{\exp(\hbar sq /k_B T)-1}
\end{equation}
is the phonon occupation number at temperature $T$, and $k_B$ is the Boltzmann
constant.

The coupling constant in (\ref{p6}) depends on
the polarization if the interaction is piezoelectric.
For longitudinal phonons, the polarization vector has Cartesian
components, expressed in terms of the spherical-coordinate angles,
\begin{equation}  \label{cpl1}
\xi_1^{\parallel}=e_1=\sin\theta \cos\phi, \quad \xi_2^{\parallel}=e_2=\sin\theta
\sin\phi,\quad \xi_3^{\parallel}=e_3=\cos\theta,
\end{equation}
where $e_j=q_j/q$. For transverse phonons, it is convenient to define the two
polarization vectors $\xi_i^{\perp1}$ and $\xi_i^{\perp2}$ to have
\begin{equation}  \label{cpl2}
\xi_1^{\perp1}=\sin\phi,  \quad \xi_2^{\perp1}=-\cos\phi, \quad \xi_3^{\perp1}=0,
\end{equation}
\begin{equation}\label{cpl4}
\xi_1^{\perp2}=-\cos \theta\cos\phi,  \quad \xi_2^{\perp2}=-\cos \theta\sin\phi,
\quad \xi_3^{\perp2}=\sin \theta.
\end{equation}

Then for longitudinal phonons, one obtains \cite{dd_ieee}
\begin{eqnarray}
w^{\parallel}&=&\frac{\pi}{\rho s V q
}M^2e^{-a^2q^2/4}\\
&&\times\, 9\sin^4\theta\cos^2\theta\sin^2\phi\cos^2\phi\sin^2
(qL\cos\theta/2).\nonumber
\end{eqnarray}
For transverse phonons, one gets
\begin{eqnarray}\nonumber
w^{\perp1}&=&\frac{\pi}{\rho s V q
}M^2e^{-a^2q^2/4}(-2\sin\theta\cos^2\theta\sin\phi\cos\phi\\
&&+\sin^3\theta\cos\phi\sin\phi)^2 \sin^2(qL\cos\theta/2),
\end{eqnarray}
\begin{eqnarray}\nonumber
w^{\perp2}&=&\frac{\pi}{\rho s V q
}M^2e^{-a^2q^2/4}(-2\sin\theta\cos\theta\cos^2\phi\\
&&+\sin\theta\cos\theta\sin^2\phi)^2 \sin^2 (qL\cos\theta/2).
\end{eqnarray}
By combining these contributions and substituting them in (\ref{W}), one can obtain the
probability of absorption of a phonon for all polarizations,
\begin{eqnarray}\label{r1}
 W_{\rm piezo}^a&=&\displaystyle\frac{M^2}{20\pi\rho s^2\hbar L^5 k^4}\frac{\exp{
 \left(-\frac{a^2 k^2}{2}\right)}}{\exp\left(\frac{\hbar s k}{k_B T}\right)-1}
 \\\nonumber
 &&\times\left\{\left(kL\right)^5+ 5 kL\left[ 2
\left(kL\right)^2 - 21\right]
\cos\left(kL\right)\right.\\
&&+\left.\nonumber 15\left[7 -
3\left(kL\right)^2\right]\sin\left(kL\right)\right\} ,
\end{eqnarray}
where
\begin{equation}
 k=\frac{\varepsilon}{\hbar s}
\end{equation} is the wave-vector of the absorbed
phonon.

Finally, the expressions for the phonon emission rates, $W^e$, can be obtained by
multiplying the above expression, (\ref{r1}), by
$(N_{th}+1)/N_{th}$.

The $\pi$ phase gate is a unitary operator which does not change the absolute values of
the probability amplitudes of a qubit in the superposition of the $|0\rangle$ and
$|1\rangle$ basis states. instead it increases the relative phase between the probability
amplitudes by  $\pi$ angle. Consequently,  superposition of $|0\rangle$ and $|1\rangle$
transforms according to
\begin{equation}
    {\Pi} \left(\alpha|0\rangle + \beta |1\rangle\right) = \alpha |0\rangle - \beta |1\rangle .
\end{equation}
Over a time interval $\tau$, the $\pi$ gate can be carried out with constant
interaction parameters,
\begin{equation}
 \varepsilon_A=0
\end{equation}
and
\begin{equation}
 \varepsilon_P = \varepsilon = \frac{\pi\hbar}{\tau}.
\end{equation}

Charge qubit dynamics during implementation of phase gates was
investigated in \cite{dd}. The relaxation dynamics is suppressed during the $\pi$ gate, because
there is no tunneling between the dots. The main quantum noise then results due to pure
dephasing. It leads to the decay of the off-diagonal qubit density matrix elements,
while keeping the diagonal density matrix elements unchanged.
The qubit density matrix can be represented in this regime as \cite{basis,Palma}
\begin{equation}\label{mm0}
\rho(t) =\left(%
\begin{array}{cc}
  \rho _{00} {(0)}& \rho _{01} {(0)}e^{ - B^2(t)+i\varepsilon t/\hbar} \\\\
  \rho _{10} {(0)}e^{ - B^2(t)-i\varepsilon t/\hbar} & \rho _{11} {(0)} \\
\end{array}%
\right),
\end{equation}
with the spectral function,
\begin{eqnarray}  \label{cd02}\nonumber
B^2(t)&=&\frac{8}{\hbar^2} {\sum\limits_{\mathbf q, \lambda} }
\frac{\left| g_{\mathbf q, \lambda}\right| ^2}{\omega _q^2}\sin ^2%
\frac{\omega _qt}2\coth \frac{\hbar \omega _q}{2 k_B T}\\
&=&\frac{V}{\hbar^2 \pi^3}\int d^3 q \sum\limits_{\lambda} \frac{\left|
g_{\mathbf q, \lambda} \right| ^2}{q^2 s^2}\sin ^2\frac{qst}2\coth \frac{\hbar q
s}{2 k_B T}.
\end{eqnarray}

For the piezoelectric  interaction, the coupling constant $g_{\mathbf q,
\lambda}$ was obtained in (\ref{p6}), and expression for the spectral function
is
\begin{eqnarray}  \label{cp1}\nonumber
B^2_{\rm piezo}(t)&=&\displaystyle \frac{M^2}{2\pi ^3\hbar \rho s^3
}\!\int_0^{\infty } q^2dq\int_0^{\pi} \sin \theta d\theta \int_0^{2 \pi} d\varphi
\\
&&\times\sum\limits_{\lambda}\frac {(\xi^{\lambda}
_1e_2e_3+\xi_2^{\lambda}e_1e_3+\xi_3^{\lambda}e_1e_2)^2}{
q^3}\exp(-a^2q^2/2){}\nonumber{}
\\
&&\times\sin^2 (q L \cos\theta)\sin ^2\displaystyle \frac{qst} 2\coth
\displaystyle\frac{\hbar q s}{2 k_B T}.
\end{eqnarray}

In summary, in this section we obtained the leading-order expressions for the semiconductor double-dot qubit density matrix in the presence of
decoherence due to piezoelectric interaction with acoustic phonons during implementation of amplitude and phase gates.

\section{Quantification of Decoherence}\label{Sec3}

Quantum information processing at the level of qubits and few-qubit registers, assumes near coherent evolution, which is at best achievable
at short to intermediate times.
Therefore attention has recently shifted from large-time system dynamics in the regime of onset of thermalization,
to almost perfectly coherent dynamics at shorter times.
Since many quantum systems proposed as
candidates for qubits for practical realizations of quantum
computing require estimation of their coherence,
quantitative characterization of decoherence is crucially important for
quantum information processing~\cite{Privman98,Kane98,Vrijen,%
Hawrylak,Alex1,Alex2,Openov,Ekert,Fedichkin,Hayashi,Barrett,Ahn,Fujisawa,%
Dzurak1,Dzurak2,Kitaev3,Shnirman,Loss,Imamoglu,Rossi,Nakamura,Tanamoto,Platzman,%
Sanders,Burkard,Bandyopadhyay,Larionov,%
Cain,Smith,Ben1,Ben2,%
qec,Steane,Bennett,Calderbank,SteanePRA,Gottesman,Knill,Kitaev,%
Kitaev2,Preskill,DiVincenzo,norm,additivity,jctn}. A single measure characterizing decoherence
is highly desirable for comparison of
different qubit designs. Besides the evaluation of single
qubit performance one also has to analyze scaling of decoherence as the register
size (the number of qubits involved) increases. Direct quantitative calculations
of decoherence of even few-qubit quantum registers are not feasible. Therefore, a
practical approach has been to explore quantitative measures of
decoherence~\cite{norm}, develop techniques to calculate such measures at least
approximately for realistic one- and two-qubit systems~\cite{dd,dd_ieee}, and
then establish scaling (additivity)~\cite{additivity,jctn}) for several-qubit
quantum systems.

In this section, we outline different approaches to define and quantify
decoherence. We argue that a measure based on a properly defined as a certain operator norm of
deviation of the density matrix from ideal, is the most appropriate for quantifying decoherence in
quantum registers.

We consider  several approaches to generally quantifying the degree
of decoherence due to interactions with environment. We first mention the approach based on the asymptotic relaxation time
scales. The entropy and idempotency-defect measures are then reviewed.
The fidelity measure of decoherence is considered next.
Finally, we introduce our
operator norm measure of decoherence. Furthermore, we discuss an
approach to eliminate the initial-state dependence of the decoherence measures.

Markovian approximation
schemes typically yield exponential approach to the limiting values of the
density matrix elements for large times \cite{Abragam,therm2,therm1}. For a
two-state system, this defines the time scales $T_1$ and $T_2$, associated,
respectively, with the approach by the diagonal (thermalization) and off-diagonal
(dephasing, decoherence) density-matrix elements to their limiting values. More
generally, for large times we approximate deviations from stationary values of
the diagonal and off-diagonal density matrix elements as
\begin{equation}
 \rho _{kk} (t) - \rho _{kk}(\infty) \propto e^{ - t/T_{kk} } ,
\end{equation}\par\noindent
\begin{equation}
 \rho _{jk} (t) \propto e^{ - t/T_{jk} }  \qquad (j \ne k) .
\end{equation}\par\noindent
The shortest time among $T_{kk}$ is often identified as $T_1$. Similarly, $T_2$
can be defined as the shortest time among $T_{n \ne m}$. These definitions yield
the characteristic times of thermalization and decoherence (dephasing).

Unfortunately the exponential behavior of the density matrix
elements in the energy basis is applicable only for large
times, whereas for quantum computing applications, the short-time behavior is
usually relevant \cite{short}. Moreover, while the energy basis is natural for
large times, the choice of the preferred basis is not obvious for short and
intermediate times \cite{short,basis}. Therefore, the time scales $T_1$ and $T_2$
have limited applicability for evaluating coherence in quantum computing.

An alternative approach is based on the calculation of the entropy \cite{Neumann} of the system,
\begin{equation}
S(t)=- {\rm Tr}\left(  \rho \ln  \rho \right),
\end{equation}\par\noindent
or the first order entropy (idempotency defect)
\cite{Kim,Zurek,Zagur},
\begin{equation}
 \label{trace}
s(t)=1 - {\rm Tr} \left( \rho ^2 \right).
\end{equation}\par\noindent
Both expressions are basis independent, have a minimum at pure states and
effectively describe the degree of the state's ``purity.'' Any deviation from a
pure state leads to the deviation from the minimal values, 0, for both measures,
\begin{equation}
S_{\,\rm pure\ state}(t)= s_{\,\rm pure\ state}(t)= 0.
\end{equation}
Unfortunately, entropy measures the deviation from pure-state evolution rather than deviation from a specific ideal evolution.

The fidelity measure, considered presently, has been widely used.
If the Hamiltonian of the system and environment is
\begin{equation}\label{f0}
  H=H_S+H_B+H_I,
\end{equation}\par\noindent
where $H_S$ is the internal system dynamics, $H_B$ gives the
evolution of environment (bath), and $H_I$ describes system-bath interaction,
then the fidelity measure \cite{Dalton,Fidelity2} can be defined as,
\begin{equation}\label{f1}
F(t)={\rm Tr}_{\,S}  \left[ \, \rho _{\rm ideal}(t) \, \rho (t) \, \right].
\end{equation}\par\noindent
Here the trace is over the system degrees of freedom, and $\rho_{\rm ideal}(t)$
represents the pure-state evolution of the system under $H_S$ only, without
interaction with the environment ($H_I=0$). In general, the Hamiltonian term
$H_S$ governing the system dynamics can be time dependent. For the sake of
simplicity throughout this review we consider constant $H_S$ over time intervals
of quantum gates, cf.\ Section~\ref{Sec2}. In this case
\begin{equation}\label{f2}
\rho _{\rm ideal}(t)= e^{-iH_S t}\rho(0)\, e^{iH_S t}.
\end{equation}
More sophisticated  scenarios with qubits evolving under time dependent $H_S$
were considered in~\cite{Solenov,Brandes,Brandes2}.

The fidelity provides a measure of decoherence in terms of the difference
between the ``real,'' environmentally influenced evolution, $\rho (t)$, and the
``ideal'' evolution, $\rho_{\rm ideal} (t)$. It will attain its maximal value, 1,
only provided $\rho (t) = \rho_{\rm ideal} (t)$. This property relies on the added assumption
the $ \rho_{\rm ideal} (t)$ remains a projection operator (pure state) for all
times $t \geq 0$.

As an simple example consider a two-level system decaying
to the ground state, when there is no internal system dynamics,
\begin{equation}
\rho _{\rm ideal} (t) =\left(
\begin{array}{cc}
0 & 0 \\
0 & 1
\end{array}
\right),
\end{equation}\par\noindent
\begin{equation}
\rho (t)=\left(
\begin{array}{cc}
1-e^{-\Gamma t} & 0 \\
0 & e^{-\Gamma t}
\end{array}
\right),
\end{equation}\par\noindent
and the fidelity is monotonic,
\begin{equation}\label{f3}
F(t)=e^{-\Gamma t}.
\end{equation}\par\noindent

Note that the requirement that $\rho_{\rm ideal}(t)$ is a pure-state (projection
operator), excludes, in particular, any $T>0$ thermalized state as the initial
system state. Consider the application of the fidelity
measure for the infinite-temperature initial state of our two level system. We
get
\begin{equation} \rho (0)=\rho _{\rm ideal}(t)=\left(
\begin{array}{cc}
1/2 & 0 \\
0 & 1/2
\end{array}
\right),
\end{equation}\par\noindent
which is not a projection operator. The spontaneous-decay density matrix is then
\begin{equation}
\rho (t)=\left(
\begin{array}{cc}
1-(e^{-\Gamma t}/2) & 0 \\
0 & e^{-\Gamma t}/2
\end{array}\right).
\end{equation}\par\noindent
The fidelity remains constant
\begin{equation}\label{f3.1}
F(t)=1/2,
\end{equation}\par\noindent
and it does not provide any information of the time dependence of the decay
process.

Let us now consider the operator norms \cite{Kato} that measure the
deviation of the system from the ideal state, to quantify the degree of
decoherence, as proposed in~\cite{norm,additivity,jctn}. Such measures do not require the initial
density matrix to be pure-state. We define the deviation according to
\begin{equation}\label{deviation}
  \sigma(t)  \equiv   \rho(t)  -   \rho_{\rm ideal} (t)  .
\end{equation}\par\noindent
We can use, for instance, the eigenvalue norm \cite{Kato},
\begin{equation}\label{n11}
\left\|\sigma \right\|_{\lambda} = {\max_i} \left| {\lambda _i } \right|,
\end{equation}\par\noindent
or the trace norm,
\begin{equation}\label{tracenorm}
\left\|   \sigma  \right\|_{{\rm Tr}}  = \sum\limits_i {\left| {\lambda _i }
\right|},
\end{equation}\par\noindent
etc., where $\lambda_i$ are the eigenvalues of the deviation operator
(\ref{deviation}).
Since density operators are Hermitian and bounded, their norms, as well the norm of the
deviation, can be always defined and evaluated by using the expressions shown, avoiding the more formal mathematical definitions.
We also note that $\left\| A \right\|=0$ implies that $A=0$.

The calculation of these norms is sometimes simplified by the observation that
$\sigma(t)$ is traceless. Specifically, for two-level systems, we get
\begin{equation} \left\|   \sigma
\right\|_{\lambda} = \sqrt {\left| {\sigma _{00} } \right|^2 + \left|
{\sigma_{01} } \right|^2 } = {1 \over 2} \left\|   \sigma  \right\|_{{\rm Tr}}.
\end{equation}\par\noindent
For our example of the two-level system undergoing spontaneous decay, the norm is
\begin{equation}
\left\|   \sigma \right\|_{\lambda} = 1 - e^{-\Gamma t} .
\end{equation}\par\noindent

The measures considered above quantify decoherence of a
system provided that its initial state is given. However, in quantum computing,
it is impractical to keep track of all the possible initial
states for each quantum register, that might be needed for implementing a
particular quantum algorithm. Furthermore, even the preparation of the initial
state can introduce additional noise. Therefore, for evaluation of
fault-tolerance (scalability), it will be necessary to obtain an upper-bound
estimate of decoherence for an arbitrary initial state.

To characterize decoherence for an arbitrary initial state, pure or mixed, we
proposed~\cite{norm} to use the maximal norm, $D$, which is determined as an
operator norm maximized over all the initial density matrices(the
worst case scenario error estimate),
\begin{equation}\label{normD}
  D(t) = \sup_{\rho (0)}\bigg(\left\| \sigma (t,\rho (0))\right\|_{\lambda}  \bigg).
\end{equation}\par\noindent

For realistic two-level systems coupled to various types of environmental modes,
the expressions of the maximal norm are surprisingly elegant and compact. They
are usually monotonic and contain no oscillations due to the internal system
dynamics. Most importantly, in the next section we will establish the {\it additivity\/}  property of the maximal norm of deviation measure.

Here we conclude by presenting the expressions for this measure for the two gates for the semiconductor double-dot system
introduced in preceding section.
The qubit error measure, $D$, was obtained from the density matrix deviation from
the ``ideal'' evolution by using the operator norm approach~\cite{norm}.
After lengthy calculations, one gets
\cite{dd} relatively simple expressions for the NOT gate,
\begin{equation}\label{Da(t)}
 D_{\rm NOT}=\frac{1 - e^{ - \Gamma \tau}}{1+e^{ -\varepsilon /  k_B T  }},
\end{equation}
and for the $\pi$ gate,
\begin{equation}\label{Dp(t)}
 D_{\pi}  = \frac{1}{2}\left[ 1 - e^{ - B^2 (\tau)}\right],
\end{equation}
where all the parameters were defined in Section~\ref{Sec2}.
A realistic ``general'' noise estimate per typical quantum-gate cycle time $\tau$,
could be taken as the
larger of these two expressions.

\section{Additivity of the Decoherence Measure}\label{Sec4}

In the study of decoherence of several-qubit systems, one has to consider the degree to
which noisy environments of different qubits are correlated \cite{Palma,additivity,JPCM}.  Furthermore, if all
constituent qubits are interacting with the same bath, then there are methods to
reduce decoherence without quantum error correction, by instead encoding the state of one logical qubit in a decoherence-free
subspace of the states of several physical qubits
\cite{Zanardi,Palma,DFS,Lidar,Wubs}. In this section, we will consider
several-qubit system and assume the ``worst case scenario,'' i.e., that
the qubits experience uncorrelated noise, and each is coupled to a separate
bath. Since analytical calculations for several qubits are impractical, we have to
find some ``additivity'' properties that will allow us to estimate the error measure for
the whole system from the error measures of the constituent qubits.
For a general class of
decoherence processes, including those occurring in semiconductor qubits
considered in Section~\ref{Sec2}, we argue that maximal deviation norm measure introduced in Section~\ref{Sec3} is additive.

The decoherence dynamics of a multiqubit system is rather complicated. The loss of quantum coherence
results also in the loss of two-particle and several-particle  entanglements in the system.
The higher order (multi-qubit) entanglements are ``encoded'' in the far off-diagonal elements of
the multi-qubit register density matrix, and therefore these quantum correlations
will decay at least as fast as the products of the decay factors for the qubits
involved, as exemplified by several explicit calculations
\cite{VPDT2,Eberly1,Storcz,Eberly2}. This observation supports the conclusion
that at  large times the \emph{rates\/} of decay of coherence of the qubits
will be additive.

However, here we seek a different result. We look for additivity property which is
valid not in the regime of the asymptotic large-time decay of quantum coherence, but for short times,
$\tau$, of quantum gate functions, when the noise level, namely the value of the
measure $D(\tau)$ for each qubit, is relatively small. In this regime, we will
establish \cite{additivity}: even for strongly entangled
qubits$\,$---$\,$which are important for the utilization of the power of quantum
computation$\,$---$\,$the error measures $D$ of the individual qubits in a
quantum register are additive. Thus, the error measure for a register made of
similar qubits, scales up linearly with their number, consistent with other
theoretical and experimental observations \cite{Dalton,Suter1,Suter2}.

Thus, to characterize decoherence for an arbitrary initial state, pure or mixed, we use
the maximal norm, $D$, which was defined (\ref{normD})  as an operator norm maximized over all the possible initial
density matrices. One can show that $0 \leq D(t) \leq 1$.  This measure of
decoherence will typically increase monotonically from zero at $t=0$, saturating
at large times at a value $D(\infty) \leq 1$. The definition of the maximal
decoherence measure $D(t)$ looks rather complicated for a general multiqubit
system. However, it can be evaluated in closed form for short times, appropriate
for quantum computing, for a single-qubit (two-state) system. We then establish
an approximate additivity that allows us to estimate $D(t)$ for several-qubit
systems as well.

The evolution of the reduced density operator of
the system (\ref{f1}) and the one for the ideal density matrix (\ref{f2}) can be
formally expressed \cite{Kitaev3,Kitaev,Kitaev2} in the
superoperator notation as
\begin{equation}\label{T1}
\rho(t)=T(t)\rho(0),
\end{equation}
\begin{equation}\label{Ti}
    \rho^{(i)}(t)=T^{(i)}(t)\rho(0),
\end{equation}
where $T$, $T^{(i)}$ are linear superoperators.
The deviation matrix
can be expressed as
\begin{equation}\label{t1}
\sigma(t)=\left[T(t)-T^{(i)}(t)\right]\rho(0).
\end{equation}

The initial density matrix can decomposed as follows,
\begin{equation}\label{mixture}
    \rho(0)=\sum_{j} p_j |\psi_j\rangle\langle\psi_j|,
\end{equation}
where $\sum_j p_j=1$ and $0 \leq p_j\leq1$. Here the wavefunction set
$|\psi_j\rangle$ is not assumed to have any orthogonality properties. Then, we
get
\begin{equation}
\sigma\left(t, \rho(0)\right)=\sum_{j} p_j
\left[T(t)-T^{(i)}(t)\right]\left|\psi_j\right\rangle\left\langle\psi_j\right|.
\end{equation}
The deviation norm can thus be bounded,
\begin{equation}\label{proj}
\|\sigma(t, \rho(0))\|_{\lambda} \; \leq \;
  \left\|  \left[T(t)-T^{(i)}(t)\right]  |\phi\rangle\langle\phi|\right\|_{\lambda}.
\end{equation}
Here $|\phi\rangle$ is defined according to
\begin{equation}\nonumber
\left\|  \left[T-T^{(i)}\right]  |\phi\rangle\langle\phi|\right\|_{\lambda}=
\max_j\left\| \left[T-T^{(i)}\right]
|\psi_j\rangle\langle\psi_j|\right\|_{\lambda}.
\end{equation}
For any initial density operator which is a statistical
mixture, one can always find a density operator which is pure-state,
$|\phi\rangle\langle\phi|$, such that $\|\sigma(t,
\rho(0))\|_{\lambda}\leq\|\sigma(t, |\phi\rangle\langle\phi|)\|_{\lambda}$.
Therefore, evaluation of the supremum over the initial density operators in order
to find $D(t)$, see (\ref{normD}), can be done over only pure-state density
operators, $\rho (0)$.

Consider briefly strategies of evaluating $D(t)$ for a single qubit. We can
parameterize $\rho(0)$ as
\begin{equation}\label{parametriazation1}
  \rho(0)=U \left(
\begin{array}{cc}
  P & 0 \\
  0 & 1-P \\
\end{array}
\right)U^{\dagger},
\end{equation}
where $0\leq P \leq 1$, and $U$ is an arbitrary $2 \times 2$ unitary matrix,
\begin{equation}
U=\left(
\begin{array}{cc}
  e^{i(\alpha+\gamma)}\cos\theta & e^{i(\alpha-\gamma)}\sin\theta
\\
  -e^{i(\gamma-\alpha)}\sin\theta & e^{-
i(\alpha+\gamma)}\cos\theta \\
\end{array}\right).
\end{equation}
Then, one should find a supremum of the norm of deviation (\ref{n11}) over all
the possible real parameters $P$, $\alpha$, $\gamma$ and $\theta$. As shown
above, it suffices to consider the density operator in the form of a projector
and put $P=1$. Thus, one should search for the maximum over the remaining three
real parameters $\alpha$, $\gamma$ and $\theta$.

Another parameterization of the pure-state density operators,
$\rho(0)=|\phi\rangle\langle\phi|$, is to express an arbitrary wave function
$|\phi\rangle=\sum_j (a_j+i b_j)|j\rangle$ in some convenient orthonormal basis
$|j\rangle$, where $j=1,\ldots,N$. For a two-level system,
\begin{equation}\label{parametrization2}
    \rho(0)=\left(%
\begin{array}{cc}
  a_1^2+b_1^2 &  (a_1+i b_1)(a_2-i b_2) \\
 (a_1-i b_1)(a_2+i b_2) &  a_2^2+b_2^2 \\
\end{array}%
\right),
\end{equation}
where the four real parameters $a_{1,2},b_{1,2}$ satisfy
$a_1^2+b_1^2+a_2^2+b_2^2=1$, so that the maximization is again over three
independent real numbers. The final expressions (\ref{Da(t)}) and (\ref{Dp(t)})
for $D(t)$, for our selected single-qubit systems considered in Section
\ref{Sec2}, are actually quite compact and tractable.

In quantum computing, the error rates can be significantly reduced by using
several physical qubits to encode each logical qubit \cite{Zanardi,DFS,Lidar}.
Therefore, even before active quantum error correction is incorporated
\cite{Ben1,Ben2,qec,Steane,Bennett,Calderbank,SteanePRA,Gottesman,Knill},
evaluation of decoherence of several qubits is an important, but formidable task.
Here our aim is to prove the approximate additivity of $D_q(t)$, including the
case of the initially  \emph{entangled\/} qubits, labeled by $q$, whose
dynamics is governed by
\begin{equation}
  H=\sum_q H_q=\sum_q \left(H_{Sq}+H_{Bq}+H_{Iq}\right),
\end{equation}
where $H_{Sq}$ is the Hamiltonian of the $q$th qubit itself, $H_{Bq}$ is the
Hamiltonian of the environment of the $q$th qubit, and $H_{Iq}$ is corresponding
qubit-environment interaction. We consider a more
complicated (for actual evaluation) diamond norm \cite{Kitaev3,Kitaev,Kitaev2},
as an auxiliary quantity used to establish the additivity of the more
easily calculable operator norm $D(t)$.

The establishment of the upper-bound estimate for the maximal deviation norm of a
multiqubit system, involves several steps. We first derive a bound for this norm in
terms of the diamond norm.  Actually, for single qubits,
in several models the diamond norm can be expressed via the corresponding maximal
deviation norm. At the same time, the diamond norm for the whole quantum system
is bounded by sum of the norms of the constituent qubits by using a certain specific
stability property of the diamond norm, $K(t)$. This norm is defined as
\begin{equation}\label{supernormK}
K(t) =\|T- T^{(i)}\|_{\diamond}=\sup_{\varrho} \|
\{[T-T^{(i)}]{\raise2pt\hbox{$\scriptscriptstyle{\otimes}$}} I\} {\varrho}
\|_{\rm Tr}.
\end{equation}
The superoperators $T$, $T^{(i)}$ characterize the actual and ideal evolutions
according to (\ref{T1}), (\ref{Ti}). Here $I$ is the identity superoperator in a
Hilbert space $G$ whose dimension is the same as that of the corresponding space
of the superoperators $T$ and $T^{(i)}$, and $\varrho$ is an arbitrary density
operator in the product space of twice the number of qubits.

The diamond norm has an important stability property, proved in
\cite{Kitaev3,Kitaev,Kitaev2},
\begin{equation}\label{stability}
\|B_1 {\raise2pt\hbox{$\scriptscriptstyle{\otimes}$}}
B_2\|_{\diamond}=\|B_1\|_{\diamond} \|B_2\|_{\diamond}.
\end{equation}
Note that (\ref{stability}) is a property of the superoperators rather than that
of the operators.

Consider a composite system consisting of two subsystems $S_1$, $S_2$, with
the noninteracting Hamiltonian
\begin{equation}
H_{S_1S_2}=H_{S_1}+H_{S_2}.
\end{equation}
The evolution superoperator of the system will be
\begin{equation}
 T_{S_1S_2}=T_{S_1}{\raise2pt\hbox{$\scriptscriptstyle{\otimes}$}}
T_{S_2},
\end{equation}
and the ideal one
\begin{equation}
T_{S_1S_2}^{(i)}=T_{S_1}^{(i)}{\raise2pt\hbox{$\scriptscriptstyle{\otimes}$}}
T_{S_2}^{(i)}.
\end{equation}
The diamond measure for the system can be expressed as
\begin{eqnarray}
&&K_{S_1S_2}^{\vphantom{(i)}}=\|T_{S_1S_2}^{\vphantom{(i)}} -
T_{S_1S_2}^{(i)}\|_{\diamond}=
\|(T_{S_1}^{\vphantom{(i)}}-T_{S_1}^{(i)}){\raise2pt\hbox{$\scriptscriptstyle{\otimes}$}}
T_{S_2}^{\vphantom{(i)}}+T_{S_1}^{(i)}{\raise2pt\hbox{$\scriptscriptstyle{\otimes}$}}
(T_{S_2}^{\vphantom{(i)}}-T_{S_2}^{(i)})\|_{\diamond}\nonumber\\
&&\leq\|(T_{S_1}^{\vphantom{(i)}}-T_{S_1}^{(i)}){\raise2pt\hbox{$\scriptscriptstyle{\otimes}$}}
T_{S_2}^{\vphantom{(i)}}\|_{\diamond}+\|T_{S_1}^{(i)}{\raise2pt\hbox{$\scriptscriptstyle{\otimes}$}}
(T_{S_2}^{\vphantom{(i)}}-T_{S_2}^{(i)})\|_{\diamond} . \label{justbelow}
\end{eqnarray}
By using the stability property (\ref{stability}), we get
\begin{eqnarray}
K_{S_1S_2}^{\vphantom{(i)}}\leq\|(T_{S_1}^{\vphantom{(i)}}-T_{S_1}^{(i)}){\raise2pt\hbox{$\scriptscriptstyle{\otimes}$}}
T_{S_2}^{\vphantom{(i)}}\|_{\diamond}+\|T_{S_1}^{(i)}{\raise2pt\hbox{$\scriptscriptstyle{\otimes}$}}
(T_{S_2}^{\vphantom{(i)}}-T_{S_2}^{(i)})\|_{\diamond}=\cr
\|T_{S_1}^{\vphantom{(i)}}-T_{S_1}^{(i)}\|_{\diamond}\|
T_{S_2}^{\vphantom{(i)}}\|_{\diamond}
+\|T_{S_1}^{(i)}\|_{\diamond}\|T_{S_2}^{\vphantom{(i)}}-
T_{S_2}^{(i)}\|_{\diamond}=\nonumber\\
\|T_{S_1}^{\vphantom{(i)}}-T_{S_1}^{(i)}\|_{\diamond}+\|T_{S_2}^{\vphantom{(i)}}-
T_{S_2}^{(i)}\|_{\diamond}= K_{S_1}^{\vphantom{(i)}}+K_{S_2}^{\vphantom{(i)}}.
\end{eqnarray}

The inequality
\begin{equation}\label{Kbound}
 K\le \sum_q K_{q},
\end{equation}
for the diamond norm $K(t)$ has thus been obtained. Let us emphasize that the
subsystems can be initially entangled. This property is particularly useful for
quantum computing, the power of which is based on qubit entanglement. However,
even in the simplest case of the diamond norm of one qubit, the calculations are
extremely cumbersome. Therefore, the use of the measure $D(t)$ is preferable for
actual calculations.

For short times, of quantum gate functions, we can use (\ref{Kbound}) as an
approximate inequality for order of magnitude estimates of decoherence measures,
even when the qubits are interacting. Indeed, for short times, the interaction
effects will not modify the quantities entering both sides significantly. The key
point is that while the interaction effects are small, this inequality can be
used for {\it strongly entangled\/} qubits.

The two deviation-operator norms considered are related by the following
inequality
\begin{equation}\label{a1}
\left\|   \sigma  \right\|_{{\lambda}}\leq\frac 1 2 \left\| \sigma \right\|_{{\rm
Tr}}\leq 1.
\end{equation}
Here the left-hand side follows from
\begin{equation}
\rm{Tr} \,\sigma =\sum_j\lambda_j =0.
\end{equation}
Therefore the $\ell$th eigenvalue of the deviation operator $\sigma$
 that has the maximum absolute value, $\lambda_\ell=\lambda_{\rm{max}}$, can be
expressed as \begin{equation}\lambda_{\ell}=-\sum_{j\neq
\ell}\lambda_j.\end{equation} Thus, we have
\begin{equation}\label{a2}
 \left\|   \sigma  \right\|_{{\lambda}}=\frac 1 2\left(2 |\lambda_\ell|\right)
 \leq
 \frac 1 2\left(|\lambda_\ell|+\sum_{j\neq \ell}|\lambda_j|\right)=
\frac 1 2\left(\sum_j|\lambda_j|\right)=\frac 1 2\left\|\sigma \right\|_{\rm Tr}.
\end{equation}
The right-hand side of (\ref{a1}) then also follows, because any density matrix
has trace norm 1,
\begin{equation}\label{a3}
\|   \sigma  \|_{{\rm Tr}} = \|   \rho-\rho^{(i)} \|_{{\rm Tr}}\leq \|   \rho
\|_{{\rm Tr}}+ \|\rho^{(i)} \|_{{\rm Tr}}=2.
\end{equation}

From the relation (\ref{a3}) it follows that
\begin{equation}\label{prop}
 K(t)\le 2.
\end{equation}
 By taking the supremum of both sides of the relation (\ref{a2}) we get
\begin{equation}\label{prop1}
 D(t)=\sup_{\rho(0)}\left\|   \sigma  \right\|_{{\lambda}}\le
 \frac 12 \sup_{\rho(0)}\left\|   \sigma  \right\|_{\rm Tr}
 \le\frac 12 K(t),
\end{equation}
where the last step involves technical derivation details \cite{additivity} not
reproduced here. In fact, for a single qubit, calculations for typical qubit models
\cite{additivity} give
\begin{equation}
 D_q(t)={\frac 1 2} K_q(t).
\end{equation}
Since $D$ is generally bounded by (or equal to) $K/2$, it follows that the
multiqubit norm $D$ is approximately bounded from above by the sum of the
single-qubit norms even for the \emph{initially entangled\/} qubits,
\begin{equation}\label{DN1}
    D(t) \le \frac 12 K(t) \le \frac 12 \sum_q K_{q}(t)= \sum_q D_{q}(t),
\end{equation}
where $q$ labels the qubits.

For specific models of decoherence of the type encountered in Section~\ref{Sec2}, as well as those formulated for general studies of short-time
decoherence \cite{norm}, a stronger property has been demonstrated by deriving additional bounds
not reviewed here
\cite{additivity}, namely that the noise measures are actually equal, for low
levels of noise,
\begin{equation}\label{DN4-b}
    D(t)=\sum_q D_{q}(t)+o\left(\sum_q
D_{q}(t)\right).
\end{equation}

Thus, in this section we considered the maximal operator norm suitable for
evaluation of decoherence for a quantum register consisting of qubits immersed in
noisy environments. We established the approximate additivity property of this measure of
decoherence for multi-qubit registers at short times, for which the level of
quantum noise is low, and the qubit-qubit interaction effects are small, but
without any limitation on the initial entanglement of the qubit register.

In conclusion, we surveyed the theory of evaluation of quantum noise effects for
quantum registers. Maximal deviation norm was proposed for error estimation and
its expressions were presented for a realistic model of semiconductor double-dot qubit interacting with acoustic phonons.
Maximal deviation norm
has a unique additivity property which facilitates error rate estimation for
several-qubit registers.

\section*{Acknowledgments}
We are grateful to A.~Fedorov, D.~Mozyrsky, D.~Solenov, I.~Vagner, and D.~Tolkunov for
collaborations and instructive discussions. This research was supported by the
National Science Foundation, grant DMR-0121146.

\end{document}